# Universal visible emitters in nanoscale integrated photonics

G. Spektor[1,2], D. Carlson[1,3], Z. Newman[1,3], J.L. Skarda[4], N. Sapra[4], L. Su[4], S. Jammi[1,2], A. R. Ferdinand[1,2], A. Agrawal[6], J. Vučković[4], and S. B. Papp[1,2]

[1]*Time and Frequency Division, National Institute of Standards and Technology, Boulder, CO 80305 USA*
[2]*Department of Physics, University of Colorado, Boulder, CO 80309, USA*
[3]*Octave Photonics, LLC., Boulder, CO USA 80305 USA*
[4]*Department of Electrical Engineering and Ginzton Laboratory, Stanford University, Stanford, 94305, California, USA*
[5]*Vector Atomic, Inc., Pleasanton, CA 94566 USA*
[6]*Microsystems and Nanotechnology Division, National Institute of Standards and Technology, Gaithersburg, MD 20899, USA*
grisha.spektor@gmail.com

**Abstract:** Visible wavelengths of light control the quantum matter of atoms and molecules and are foundational for quantum technologies, including computers, sensors, and clocks. The development of visible integrated photonics opens the possibility for scalable circuits with complex functionalities, advancing both the scientific and technological frontiers. We experimentally demonstrate an inverse design approach based on superposition of guided-mode sources, allowing the generation and full control of free-space radiation directly from within a single 150 nm layer $Ta_2O_5$, showing low loss across visible and near-infrared spectra. We generate diverging circularly-polarized beams at the challenging 461 nm wavelength that can be directly used for magneto-optical traps of strontium atoms, constituting a fundamental building block for a range of atomic-physics-based quantum technologies. Our generated topological vortex beams and spatially-varying polarization emitters could open unexplored light-matter interaction pathways, enabling a broad new photonic-atomic paradigm. Our platform highlights the generalizability of nanoscale devices for visible-laser emission and will be critical for scaling quantum technologies.

Addressing matter with visible lasers enables a wide range of applications, including facilitating quantum computer technology[1–3], conveying information to the human brain[4,5], sensing various biological markers[6,7], realizing exceptional measurement precision and quantum sensing[8], and facilitating discoveries at the frontier of physical science[9–13]. Photonic-integrated circuits (PICs) of extended waveguide structures are the medium to realize complex systems for myriad applications, especially in the quest for portability[9,12]. Novel advances in nanoscale photonics will open scalable opportunities to probe, manipulate, and address various types of matter with visible laser light.

To motivate integrated photonics development, we take ensembles of atoms as an essential and widespread system at the frontier of quantum information[14] and quantum sensing[15,16]. In order to control the relative atomic motion of the ensemble and manipulate its spin to prepare and detect functional quantum states, current experiments rely on complex, free-space laser beam systems implemented with bulk optics. In particular, the extraordinarily long radiative lifetimes of atomic transitions in alkaline-earth metals such as strontium (Sr) enable ultraprecise laser frequency stabilization. This, in turn, facilitates the operation of optical-frequency clocks that potentially exceed the performance of cesium clocks used to realize the SI definition of the second[17].

Beyond the development of new physical standards, the extreme measurement precision of Sr clocks[18,19], counted at approximately 1 part in $10^{18}$, enables the discovery of novel physics[20], gravitational wave detection[21], relativistic geodesy[22,23] and dark matter searches[24]. Frequency-comb metrology applications, which are integral tools in atomic physics, have recently become enabled by integrated photonics in the near-infrared to the visible spectra[25–27]. The generation of arbitrarily-complex free space laser fields through integrated photonics, using scalable manufacturing, can unlock unprecedented opportunities with optical clocks, including quantum sensors[28], higher precision GPS clocks in space[29,30], precision geodesy[31], pristine synchronization of distributed systems[32,33], and more. Innovation in the control of visible laser beams relies on expanding our knowledge of nanoscale photonics, towards the goal of arbitrarily complex and scalable systems[34,35], given the requirement for deep subwavelength structures for use with blue and UV wavelengths. Moreover, this application space requires control of the laser field far outside the near-field range[36], increasing the demand for on-chip sources.

Given the importance of generating free-space laser beams, various approaches have been developed. With the specific aims and requirements of interfacing lasers to quantum systems, recent advances with nanoscale patterning have generated large-area[37,38] and focused[39] free-space laser beams, albeit with the emitted polarization typically constrained to be parallel to the grating lines. A combined approach of PICs with metasurfaces to add functionality has been recently explored[40], based on a non-trivial cascade of substrate bonding individually fabricated devices. The closely related polarization-diversity coupling gratings aim to couple light *into* PICs regardless of their polarization[41–50]. However, their design has consisted of involved simulation campaigns to optimize a-priori selected structure arrays and tailor them for a given wavelength. Finally, the inverse-design approach[51], relying on advanced optimization algorithms to implement physically intuitive ideas through the generation of physically unintuitive structures, has recently transformed the field of photonics[52,53]. This approach offers general solutions to many problems, including linear[54–58] and nonlinear[59–61] integrated elements, light-trapping devices[62], free-space-to-free-space metasurfaces[63–66], and free-space-to PIC coupling gratings[67–69]. Nevertheless, a generalized platform for PIC-to-free-space beam delivery with complete control over all properties of emitted light has yet to be realized.

In this work, we leverage nanoscale devices to demonstrate a versatile PIC platform for laser emission across the visible and near-infrared spectral ranges. We experimentally demonstrate that leveraging coherent superpositions of waveguide modes (Fig. 1 a) and utilizing an inverse design[51] approach enables the generation of arbitrarily shaped, oriented, and polarized laser beams directly from a single 150 nm thin PIC layer (Fig 1 c-f). We perform the design fabrication and characterization experiments at the challenging 461 nm wavelength, which is critical for laser cooling and trapping of strontium gases, thus laying the foundation for a generalized atomic-photonic platform. We introduce the tantala ($Ta_2O_5$) material system, fully integrated on an oxidized silicon wafer, and supports record low-loss[70,71] down to 3±1 dB/m across the telecommunications C-band. Tantala is a CMOS-compatible material[72] that has recently emerged for integrated nonlinear photonics[71,73–76] and, as such, holds high potential for becoming a scalable basis for photonic-atomic circuitry. The generation of highly divergent beams allows for a small footprint (<10 μm²) structure that facilitate high robustness to fabrication tolerances and enable tight placement of a virtually unlimited number of various sources, eliminating bulky optics typically used in atomic physics setups. Our platform provides a clear pathway to miniaturize existing quantum technologies and export them from their laboratory-bound environments. Moreover, the demonstrated generation of topological sources such as vortices and beams with spatially-varying polarization opens new directions towards unexplored light-matter interaction, opening and enabling a broad new photonic-atomic paradigm.

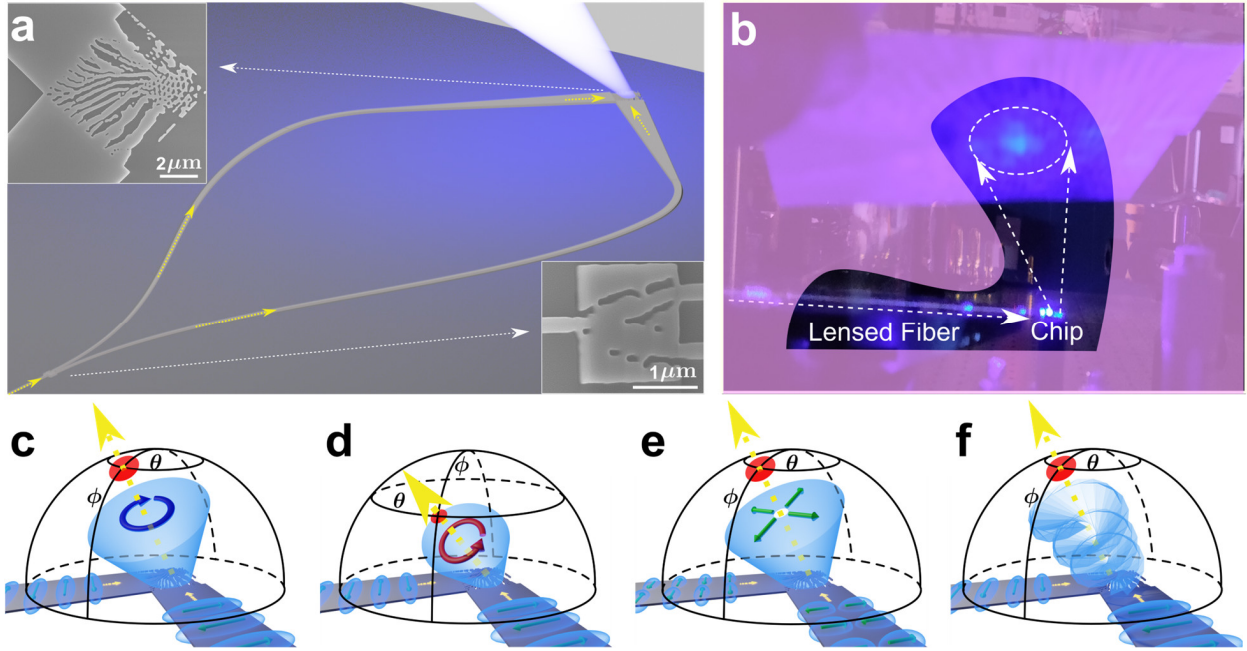

**Figure 1.** Universal interface for free-space beam emission, based on nanoscale structures. (a) Concept schematic of a guided mode split into two modes propagating towards an inverse-designed two-dimensional structure, generating a diverging free-space beam with structurally-defined arbitrary polarization. Scanning Electron Microscope (SEM, top left inset) image of a sample with inverse-designed beam source structure determining the beam spatial properties in an air-clad, 150 nm thin tantala-on-oxide platform. The relative phase and amplitude of the two waveguide arms are defined by an inverse-designed phase splitter (SEM bottom right inset) and determine the polarization state of the generated beam. (b) Photograph of a blue, free-space, polarized beam with 461 nm wavelength originating from the fiber-coupled chip. (c-f) Schematic representation of the demonstrated controlled degrees of freedom. The yellow dashed arrow indicates the propagation direction pointing orientation ($\theta$, $\varphi$). The beam cone size and red spot diameter indicate the divergence angle. Polarization: blue arrow - left circular polarization (c), red arrow - right circular polarization (d), spatially varying polarization (e) in the form of radial polarization generated by scattering the singularity carried by two perpendicularly propagating first-order modes, and a spatially varying phase in the form of a phase vortex (f).

The basis behind on-chip polarization control with a nanoscale dielectric structure considers a spatially uniform elliptical polarization state $\hat{e}$ as a linear superposition of two orthogonal polarization vectors,

$$\hat{e} = a\hat{x} + b \cdot \exp i\phi\, \hat{y} \qquad (1)$$

In a typical structure for free-space laser emission outcoupling, a fundamental waveguide mode is guided towards a grating. The mode is scattered into free space, with the resulting polarization typically lying parallel to the grating lines. The spatial-field distribution is achieved using grating apodization since the amplitude and phase of the guided mode change with propagation and partial scattering along the grating. The apodization is typically achieved via maximizing a field distribution target by selecting teeth from a pre-simulated grating library, assuming periodic structures and exponential propagation decay inside the gratings.

To achieve two-dimensional beam control, it is, in principle, possible to add an orthogonally oriented waveguide arriving at the outcoupling grating. Superimposing the outcoupling port with an orthogonally oriented grating would combine the orthogonally-polarized guided modes as they scatter into free-space beams. Then, by controlling the relative amplitude and phase of the guided modes, one could control the polarization in the output according to equation (1). This scenario, conceptually reversing the polarization-diversity insertion couplers, is achievable by apodized gratings (see SI). However, the generation of highly divergent beams propagating at steep angles in two dimensions, necessary to support the broadest range of technologies, by use of apodization of orthogonal gratings is a non-trivial grating-design task. Moreover, achieving optimal coupling efficiencies is challenging with such approaches, and generalizing to more involved functionalities is strongly limited by the initial model and the employed assumptions.

The alternative strategy we take here is a direct optimization of the structure to maximize a chosen output beam target via inverse design. Inverse design is an optimization approach that achieves local optimal values for a given objective function over vast parameter spaces. Once a physical framework is set to provide enough degrees of freedom, including, specifically, nanoscale feature control in photonics platforms, the operation of inverse design can result in devices with optimal performance[53]. We show here that the generalizability of inverse design is ideal for matching the requirements of quantum technologies. A general electromagnetic optimization problem can be expressed[51] in the form,

$$\max_{p} f_{\text{obj}}\left(\boldsymbol{E}(\epsilon(\boldsymbol{p}))\right)$$
$$\text{Subject to: } \boldsymbol{p} \in S_{\text{fab}}$$
(2)

where $f_{\text{obj}}$ is an objective function to be maximized, $\boldsymbol{E}$ is the electric field distribution, $\epsilon$ is the distribution of the electric permittivity, and $\boldsymbol{p}$ is some parametrization of the permittivity containing the fabrication constraints $S_{fab}$[51]. We specify the desired beam with its target electric and magnetic vector fields $\boldsymbol{E}_{\text{target}}, \boldsymbol{H}_{\text{target}}$, which can be defined analytically, and by setting the orthogonal source input configuration (Fig. 1 a, b), we maximize the overlap objective given by,

$$f_{\text{obj}} = 0.25 \cdot \left[\frac{\int \boldsymbol{E}(\epsilon) \times \boldsymbol{H}^{*}_{\text{target}} \cdot dS}{N} + \frac{\int \boldsymbol{E}^{*}_{\text{target}} \times \boldsymbol{H}(\epsilon) \cdot dS}{N^{*}}\right]$$
(3)

Where $\boldsymbol{E}(\epsilon), \boldsymbol{H}(\epsilon)$ are the permittivity-dependent electric and magnetic vector fields, $N = 0.5 \cdot \int \boldsymbol{E}_{\text{target}} \times \boldsymbol{H}^{*}_{\text{target}} \cdot dS$, and the integral is calculated over the region of interest of sufficiently non-vanishing fields.

Circular polarization states are central to preparing optimized ensemble spin states in quantum applications. Therefore, as the first example, we explore the generation of circularly polarized beams propagating at $\phi = 210^o$ azimuthal and $\theta = 20^o$ polar angles with a high divergence diameter of $20^o$. We explicitly formulate this objective beam configuration as an overlap with a circularly polarized Gaussian beam with the above properties and set two orthogonally propagating guided fundamental TE modes with a $\pi/2$ relative phase as simultaneous inputs. To demonstrate low footprint devices, we define a beam having a 1.5 µm waist located 300 nm above the top surface of the device layer, where we evaluate the overlap integral. In addition, this allows us to minimize optimization and simulation times and generate devices with a tiny footprint of 6 µm². To characterize the properties of the beams generated by the structures, we first conduct finite difference time domain (FDTD) simulations. We calculate the far-field intensity distribution and show the beam's angular propagation and divergence properties (Fig. 2 b). We compute the Stokes parameters of the resulting fields and plot them as a point cloud on the Poincaré sphere to obtain the polarization distribution information. Figure 2 d shows the point cloud lying at the bottom of the Poincaré sphere with the intensity weighted average represented by the red dot. To characterize the polarization distribution within the beam, we calculate the pointwise third Stokes parameter $S_3 = -2 \cdot \text{Im}(E_x E_y^*)$, representing the circular polarization purity, and verify (Fig. 2 c) that the beam is circularly polarized. We show that the obtained structures (Fig. 1 b) have the desired performance (Fig. 2), exhibiting simulated outcoupling efficiencies of up to 50% (see SI).

To experimentally characterize our nanophotonic design principle for visible laser emission, we fabricate devices using electron-beam lithography with tantala films[71], which are deposited by ion-beam sputtering onto an oxidized silicon wafer. To control the phase between the two guided sources, we inverse-design phase beam splitters that receive a single input and can assign desired phase and amplitude to the two outputs (Fig. 1 a and Fig. 4 b). For the optical-field analysis, we construct a polarization microscope (Fig. 2 e) starting with a high numerical aperture (NA=0.9) objective, allowing the collection of highly divergent beams propagating at steep angles. The light from the infinity-corrected objective passes through a rotating quarter waveplate and a linear polarizer, followed by a tube lens and a transforming lens projecting the far-field onto a detector. The system provides the angular far-field intensity distribution and pixel-wise polarization information (Fig. 2 f-h), providing a complete detailed characterization of the beams (See SI).

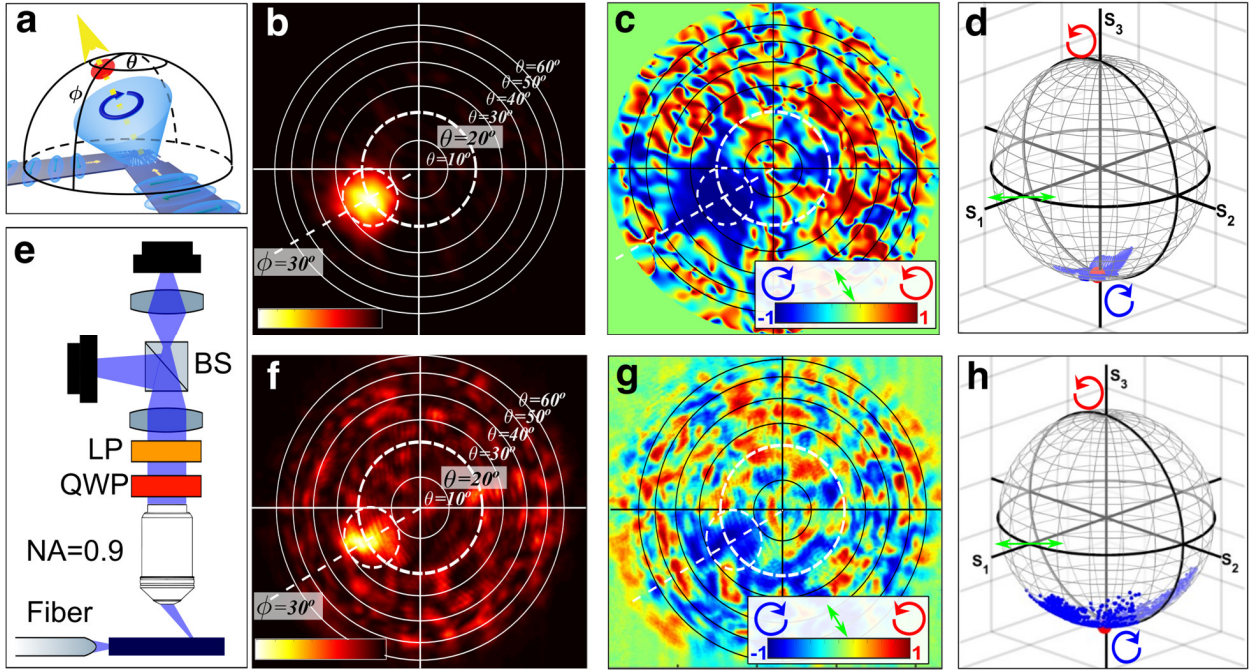

**Figure 2.** Experimental demonstration. (a) schematic of the diverging left-circularly-polarized (blue circular arrow) beam. (b) Far-field intensity angular distribution modeling of a structure generating a diverging beam at $\lambda = 461$ nm propagating at ($\theta = 20°, \phi = 30°$) with $10°$ divergence radius (white dashed line encircling the beam). The circular purity spatial distribution ($S_3$ stokes parameter) (c) and the Poincaré sphere (d) both show that the beam is circularly polarized. (c) Shows the polarization distribution within the divergence radius of the beam concentrated around $S_3 = -1$, at the south pole of the Poincaré sphere (d). The point cloud on the Poincaré sphere contains points within the beam with normalized intensity values greater than 0.1. The red dot represents the intensity-averaged polarization located at the south pole. (e) Spatial polarimeter setup schematic. Light at 461nm is coupled to a tantala-on-oxide photonic chip. The highly divergent beams are collected by an infinity-corrected objective (numerical aperture 0.9). The light from the objective passes through a rotating quarter waveplate (red) followed by a linear polarizer (orange). Following a tube lens, the light is split into a near-field image (left camera) and into a lens that projects the far-field of the illumination onto the top camera. (f) Experimental measurement of the far-field intensity distribution and (g) circular polarization purity as well as the Poincaré sphere showing the polarization distribution within the beam having a good agreement with the modeled results.

As shown above, the direct optimization design approach can utilize the degrees of freedom made available by a proper physical system to result in functioning devices. However, the convergence to local maxima can hinder their performance and polarization quality and severely increase simulation time. To obtain post-optimization experimental control over the emitted polarization with a single grating device, we show that it is possible to re-cast our optimization problem by aggregation of several objectives,

$$f_{\text{obj}}(\boldsymbol{p}) = \sum_i f_i(\boldsymbol{p})^2 \qquad (2)$$

where $f_i(\boldsymbol{p})$ are the individual optimization objectives. We demonstrate the approach by designing a grating structure allowing for a controllable polarization state. Our objective function now becomes

$$f(\boldsymbol{p}) = f_{\text{horiz}}(\boldsymbol{p})^2 + f_{\text{vert}}(\boldsymbol{p})^2 \qquad (5)$$

where $f_{\text{horiz}}(\boldsymbol{p})$ and $f_{\text{vert}}(\boldsymbol{p})$ are two overlap integrals with two different orthogonally polarized Gaussian beams sharing the same spatial properties (divergence, propagation orientation). The objectives are maximized jointly, where each objective is now achieved by a single guided mode source. This is contrary to the previous case where both sources are simultaneously active to maximize the single objective. As depicted in Figure 3 c, a $20^o$ diameter diverging beam propagating at $(\phi = 30^o, \theta = 20^o)$ is obtained, and by changing the phase and amplitude of the two guided input sources, we achieve different polarization states while maintaining the spatial beam properties using the same jointly optimized grating structure.

A critical tool in atomic physics is the magneto-optical trap (MOT)[77]. The optical part of the MOT consists of 3 pairs of counter-propagating, circularly-polarized beams that are sent at a $\theta = 45^o$ inclination, evenly distributed along the azimuthal angle. These 6 beams all meet at a common center where the atoms are eventually trapped. We show that our platform enables a photonic-based MOT by generating a circularly polarized beam (Fig. 3 b) propagating at $(\phi = 30^o, \theta = 45^o)$. The experimentally obtained beam shows a slight distortion due to interfering reflections from the Si substrate, which can be readily resolved in the future by etching windows in the substrate or as described in the SI. This key demonstration at the strontium MOT wavelength of 461 nm unlocks the possibility of generating a fully integrated photonic-atomic trap for ultracold Strontium atoms or various other atomic or molecular species.

To further demonstrate the versatility of our approach, we show that we are not restricted by spatially-uniform polarizations and allow the generation of spatially-varying vector fields,[78] having a nonuniform polarization distribution across the profile of the beam. For example, we show how to generate a radially-polarized diverging beam (Fig. 3 d), having a donut-shaped spatial distribution with a polarization singularity at its dark center. The beam's electric field points radially outward away from the central singularity. Such a polarization state manifests as a belt along the equator of the Poincaré sphere (Fig. 3 d). We achieve the polarization state by optimizing for a structure that accepts two first-order guided modes carrying a polarization singularity and outputs a radially-polarized beam. We implement a mode converter that converts the fundamental mode into the first-order mode with 95% conversion efficiency. Two resulting first-order modes are guided towards the grating in the same manner as previous geometry (see SI). Such beams are useful in microscopy applications generating the smallest focal spots below the Rayleigh limit[79,80] and could fuel future optical tweezer-based atomic applications.

Finally, to demonstrate the control over the spatial phase distribution of the generated beams, we create a beam carrying a phase singularity (Fig. 3 e). In contrast to the radially polarized beam, the phase vortex has a spatially constant polarization (vertical in this case) and zero intensity in its center, around which the phase of the field accumulates $2\pi$ phase, forming an optical vortex[81,82] carrying a unit of optical orbital angular momentum. Such beams have applications[82] ranging from optical communications[83] to tweezing[84] and quantum states[85] and have demonstrated various uses in atomic physics[86,87], including mechanical manipulations such as rotation, trapping, and routing[88] as well as generation of macroscopic matter waves[89], and even transferring angular momentum to electrons[90]. Recent work[91] demonstrated the generation of vortex beams in the telecom wavelengths; nevertheless, our results at more than three-fold shorter wavelength (461 nm) open the door for vortex-beam atomic-photonic applications. Our platform facilitates the integration of many such topological sources at short visible wavelengths suitable for interactions with atoms, opening and enabling a broad new photonic-atomic paradigm.

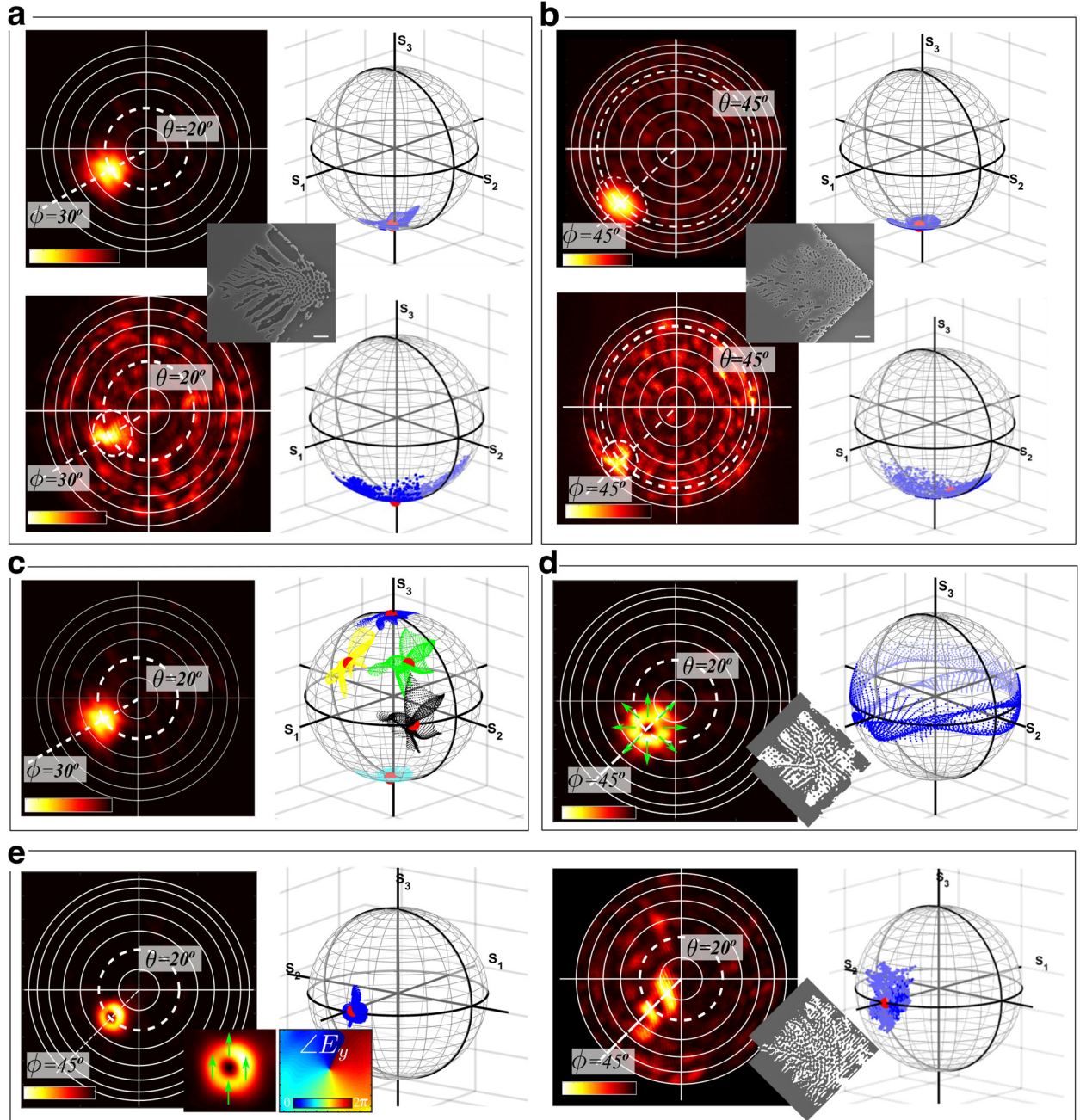

**Figure 3.** Demonstrating degrees of freedom of beam source control. (a, b) Controlling beam propagation properties while maintaining circular polarization. Simulated (top) and experimental (bottom) angular far-field beam intensity profiles of a structure generating for a $\lambda = 461$ nm beam propagating at ($\theta = 20^o, \phi = 30^o$) with $10^o$ divergence radius (a) and ($\theta = 45^o, \phi = 45^o$) with $15^o$ divergence radius (b), the insets are the SEM images of the generating structures. The simulated and experimental Poincaré spheres of the generated beams. Each point cloud on the sphere depicts the points within the beam with normalized

intensity greater than 0.1. The red circles represent the intensity-weighted mean polarization of the whole beam. The polar beam orientation, as well as divergence angle, are controlled while maintaining a constant azimuthal orientation and a left circular polarization located on the south pole of the Poincaré sphere. (c) Controlling polarization while maintaining the normalized beam profile. Far-field beam intensity profile simulation of a structure generating for a $\lambda = 461$ nm beam propagating at ($\theta = 20^o, \phi = 30^o$) with $10^o$ divergence radius with the same generating structure as in (a) fed with varying phase and amplitude splitters resulting in the same spatial profile but different polarizations. The Poincaré sphere shows the polarization within each beam; each point-cloud color corresponds to a different beam with a practically unaltered spatial profile but altered polarization state. The control is achieved by altering the relative phase and amplitudes of the source waveguided modes using integrated phase-shifting splitters using the same grating structure as in (a). (d) Spatially varying polarization generation. Angular far-field intensity beam profile simulation of a structure generating a $\lambda = 461$ nm radially polarized beam propagating at ($\theta = 20^o, \phi = 45^o$). The field distribution clearly shows the doughnut-shaped beam representing the polarization singularity at the center. The polarization content of the beam covers a band around the equator of the Poincaré sphere spanning all linear polarization states with a slight ellipticity manifested in the widening of the band. For full generation schematic, see SI. (e) Controlling the spatial phase distribution of a beam. Simulated (left) and experimental (right) far-field intensity distribution and Poincaré sphere of abeam having a phase singularity with a doughnut intensity distribution, while the polarization remains linear and constant within the beam, concentrated around a single point on the equator of the Poincaré sphere exhibiting a phase accumulation of $2\pi$ radians around the dark singularity (insets bottom left).

A platform providing the versatility to support diverse laser-beam emission objectives and the overall application space for visible integrated photonics requires a full palette of devices to control waveguide propagation. We introduce the desired functionalities and implement them using inverse design, fabricating the devices on the tantala-on-oxide platform. We created waveguides of varying lengths to characterize the propagation losses exhibited by the tantala waveguides (Fig. 4 a). We demonstrate that our platform exhibits <1 dB/cm propagation losses for 780 nm (0.29±0.03 dB/cm) and 510 nm (0.57±0.07 dB/cm) wavelengths and 1.45±0.06 dB/cm for the challenging 450 nm wavelength. We design a set of passive elements to guide, manage, and deliver the light within the optical chip (Fig. 4 b-e and SI). These elements allow us to split the light while controlling both amplitude and phase between two (Fig. 4 b) or more (Fig. 4 c) outputs; generate multiple on-chip sources required in most quantum applications such as magneto-optical traps; allow the routing of different colors within a single layer (Fig. 4 d-e); convert to high order guided modes, and filter undesired polarization components (see SI). The designed passive elements exhibit compact footprints and over 90% efficiencies, allowing the generation of complex circuit schemes with many accurately controlled sources in the visible, demonstrating that tantala provides a versatile platform for visible photonics.

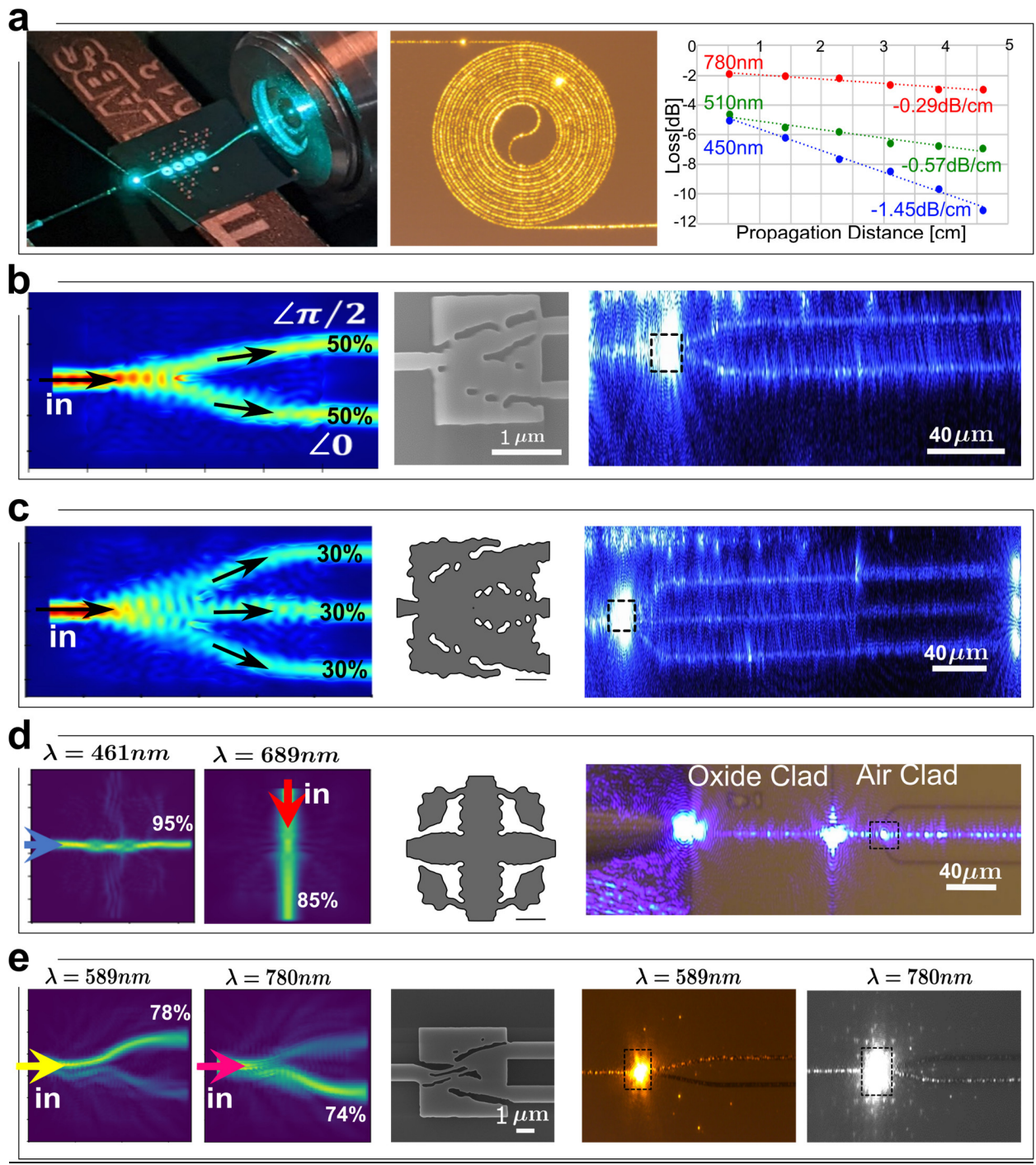

**Figure 4.** Tantala losses and light routing elements. (a) Loss characterization of tantala waveguides, (left) photo of a fiber exciting a length-scan chip at 510 nm propagating ~5 cm inside the chip, (middle) a single length scan spiral element, (right) length scan at three different wavelengths. (b) 1x2 461 nm phase-shift splitter, assigning a $\pi/2$ phase difference between its arms, (left) FDTD modeling (middle) SEM image, and (right) experimental scattering image of the splitter used to determine the free space beam polarization. (c) 1x3 461 nm splitter designed for routing light towards three blue MOT sources (left) modeling (middle) structure schematic, scale bar 500 nm, (right) experimental scattering image of the device. (d) Blue/red crossing designed for routing the blue and red MOT beams within a single tantala layer, (left) modeling (middle) design schematic (scale bar 500 nm), (right) experimental scattering image of the device, light is coupled from the left via a lensed fiber into an oxide-clad region, scattering by the air/oxide-clad interface followed by the crossing showing that no light is coupled to the crossed route with $10^{-4}$ cross-talk. (e)

wavelength division structure routing 589 nm and 780 nm light from the common input to different outputs, (left) modeling (middle) SEM of the device, (right) scattering images of the device showing different wavelength operation.

To summarize, we have demonstrated and experimentally investigated an inverse design approach based on the superposition of guided mode sources, allowing the generation and complete control of free-space beams directly from within a single layer of a photonic chip. We implemented our approach by using a single 150 nm thin $Ta_2O_5$ layer on oxide at a challenging wavelength of 461 nm, suitable for the blue magneto-optical trap of strontium atoms. The containment of all elements within a single fabrication layer means that our platform is readily scalable to enable the shaping, and delivery of a virtually unlimited number of closely spaced sources at various visible colors, thus opening new avenues for quantum technologies. We believe that our platform would serve as an enabling technology in many areas requiring truly functional and scalable integration of photonic systems and would be instrumental in commodifying fieldable quantum technologies.

## Methods

Modeling. The modeling is done using, Lumerical, Spins, and Matlab.

Fabrication. 150 nm of tantala is deposited on thermally oxidized Si substrates via ion beam sputtering. A layer of Ma-N 2404 e-beam resist is spin-coated onto the sample, patterned using electron-beam lithography, and etched in an inductively coupled plasma reactive-ion etcher using fluorine-based chemistry. After stripping the resist, the chip outlines are patterned using the optical lithography step defines the individual chip outlines, which are then diced via deep reactive ion etching. A final sulphuric-acid-based cleaning step is performed before annealing in the air for approximately 10 hours at 500° C.

Experiment. A 461 nm laser (Toptica) is delivered via a polarization-maintaining fiber (Thorlabs). The power and polarization are stabilized using a noise eater (Thorlabs NEL01A) and coupled via a custom lens fiber (Nanonics Imaging) into a chip. The light from the chip is gathered with a high NA objective (NA 0.9 CFI Plan Achromat NCG 100X). The collimated light passes through a constant polarizer and is followed by a rotating quarter waveplate (QWP), and the far-field image is collected by a camera. The polarization data is collected by rotating the QWP with a 1 degree angular step completing a full 360 degree revolution. The spatial polarization distribution is extracted by pixel-wise fitting and described in more detail in the SI.

Disclaimer: Mention of specific products or trade names is for technical and scientific information and does not constitute an endorsement by NIST.

## Acknowledgments

G.S. Acknowledges the support of the Schmidt Science postdoctoral fellowship and the Viterbi fellowship. The work is funded by NIST and the DARPA A-PhI program FA9453-19-C-0029.